\documentclass{elsarticle}

\usepackage{amsmath}
\usepackage{graphicx}

\input{tcilatex}

\begin{document}

\begin{frontmatter}
\title{A Test for the Presence of a Signal, \newline with Multiple Channels and Marked Poisson}
\author[wolf]{Wolfgang A. Rolke}
\address[wolf]{Department of Mathematics, University of Puerto Rico - Mayag\"{u}ez, Mayag\"{u}ez, PR 00681, USA, 
\newline Postal Address: PO Box 5959, Mayag\"{u}ez, PR 00681, 
\newline Tel: (787) 255-1793, Email: wolfgang@puerto-rico.net}
\author[ang]{Angel M. L\'{o}pez}
\address[ang]{Department of Physics, University of Puerto Rico - Mayag\"{u}ez, Mayag\"{u}ez, PR 00681, USA}
\begin{abstract}
   We describe a statistical hypothesis test for the presence of a signal based on the likelihood ratio statistic. We derive the 
test for a special case of interest. We study extensions of the test to cases where there are multiple channels and to marked Poisson distributions.
We show the results of a number of performance studies which indicate that the test works very well, even far out in the tails of the distribution and
with multiple channels and marked Poisson.
\end{abstract}
\end{frontmatter}\newpage

\section{Introduction}

One of the main goals of the upcoming experiments at the Large Hadron
Collider at CERN will be to make discoveries, for example of the Higgs
boson. To do so it will be necessary to make use of all available
information, that means we will need to use data from multiple channels as
well as auxiliary measurements. In this paper we will describe a test
capable of doing so, based on the likelihood ratio test statistic. The main
contribution of this paper is the study of the performance of this test.

Discoveries in high energy physics require a very small false-positive, that
is the probability of falsely claiming a discovery has to be very small.
This probability, in statistics called the type I error probability $\alpha $%
, is sometimes required to be as low as $2.87\cdot 10^{-7}$, equivalent to a 
$5\sigma $ event. The likelihood ratio test is an approximate test, and what
sample sizes are necessary for the approximation to work, especially this
far out in the tail, is a question that needed to be investigated.

\section{Likelihood Ratio Test}

The general problem of discovery is as follows: we have data $\mathbf{X}$
from a distribution with density $f(\mathbf{x};\theta )$ where $\theta $\ is
a vector of parameters with $\theta \in \Theta $ and $\Theta $ is the entire
parameter space. We wish to test the null hypothesis $H_{0}:\theta \in
\Theta _{0}$ (no signal) vs the alternative hypothesis. $H_{a}:\theta \in
\Theta _{0}^{c}$\ (some signal), where $\Theta _{0}$ is some subset of $%
\Theta $. The likelihood function is given by%
\begin{equation*}
Like(\theta |\mathbf{x})=f(\mathbf{x};\theta )
\end{equation*}%
and the likelihood ratio test statistic is defined by%
\begin{equation*}
\lambda (\mathbf{x})=\frac{\sup_{\Theta _{0}}Like(\theta |\mathbf{x})}{%
\sup_{\Theta }Like(\theta |\mathbf{x})}
\end{equation*}%
Because $Like(\theta |\mathbf{x})\geq 0$ and because the supremum in the
numerator is taken over a subset of the supremum in the denominator we have $%
0\leq \lambda (\mathbf{x})\leq 1$. The likelihood ratio test rejects the
null hypothesis if $\lambda (\mathbf{x})\leq c$, for some suitably chosen $c$%
, which in turn depends on the type I error probability $\alpha $.

How do we find $c$? There is of course a famous theorem that states that under some
mild regularity conditions, if $\theta \in \Theta _{0}$ then $L(\mathbf{x}%
)=-2\log \lambda (\mathbf{x})$ has a chi-square distribution as the sample
size $n\rightarrow \infty $. The degrees of freedom of the chi-square
distribution is the difference between the number of free parameters
specified by $\theta \in \Theta _{0}$ and the number of free parameters
specified by $\theta \in \Theta $.

A proof of this theorem is given in Stuart, Ord and Arnold \cite{Stuart et
al} and a nice discussion with examples can be found in Casella and Berger 
\cite{Casella and Berger}. Unfortunately the theorem does not apply to our case,
nevertheless as we shall see the conclusion does. 

\section{An Example: A Counting Experiment with Background, Efficiency and
Acceptance}

We begin with a very common type of situation in high energy physics
experiments. This is a search for a particle by observing a particular decay
channel. After suitably chosen cuts we find $n$ events in the signal region,
some of which may be signal events. We can model $n$ as a random variable $N$
with a Poisson distribution with rate $res+b$ where $b$ is the background
rate, $s$ the signal rate for the production of the particle, $e$ the
efficiency for observing the particular decay channel and $r $ the branching
fraction to that channel. We also have an independent measurement $y$ of the
background rate, either from data sidebands or from Monte Carlo and we can
model $y$ as a Gaussian random variable $Y$ with rate $b$ and standard
deviation $\sigma _{b}$. Finally we have an independent measurement of the
efficiency $z$, usually from Monte Carlo, and we will model $z$ as a
Gaussian random variable $Z$ with mean $e$ and standard deviation $\sigma
_{e}$. $\sigma _{b}$, $\sigma _{e}$ as well as the branching fraction $r$
are assumed to be known. So we have the following probability model:%
\begin{equation*}
N\sim Pois(res+b)\qquad Y\sim N(b,\sigma _{b})\qquad Z\sim N(e,\sigma _{e})
\end{equation*}%
In this model $s$ is the parameter of interest and $e$ and $b$ are nuisance
parameters. Now the joint density of $N$, $Y$ and $Z$ is given by

\begin{equation*}
\begin{tabular}{l}
$P(N=n,Y=y,Z=z)dydz=f(n,y,z;e,s,b)=$ \\ 
$\frac{\left( res+b\right) ^{n}}{n!}e^{-\left( res+b\right) }\frac{1}{\sqrt{%
2\pi \sigma _{b}^{2}}}e^{-\frac{1}{2}\frac{(y-b)^{2}}{\sigma _{b}^{2}}}\frac{%
1}{\sqrt{2\pi \sigma _{e}^{2}}}e^{-\frac{1}{2}\frac{(z-e)^{2}}{\sigma
_{e}^{2}}}$%
\end{tabular}%
\end{equation*}

Finding the likelihood ratio test statistic $\lambda $ means maximizing the
density above (now viewed as the likelihood) twice, once over all parameters
and then again assuming $s=0$. We find

\begin{equation*}
\begin{tabular}{l}
$L(n,y,z)=(-2)\log \left\{ \frac{\sup_{\{b,e\}}Like(0,b,e;n,y,z)}{%
\sup_{\{s,b,e\}}Like(s,b,e;n,y,z)}\right\} =$ \\ 
$2n\log (n/\widetilde{b})+2\widetilde{b}-2n+\frac{(y-\widetilde{b})^{2}}{%
\sigma _{b}^{2}}$ \\ 
where $\widetilde{b}=\frac{1}{2}\left( y-\sigma _{b}^{2}+\sqrt{(y-\sigma
_{b}^{2})^{2}+4n\sigma _{b}^{2}}\right) $%
\end{tabular}%
\end{equation*}

First we note that the test statistic does not involve $z$, the estimate of
the efficiency, nor does it involve $r$, the branching fraction. This is
true for the one channel case but will no longer hold for multiple channels,
although we will find that the test is sensitive only to the relative
efficiencies and relative branching ratios between channels, quantities
which are usually known more precisely than the absolute values.

Now from the general theory we know that $L(N,Y,Z)$ has a chi-square
distribution with $1$ degree of freedom because in the general model there
are $3$ free parameters and under the null hypothesis there are $2$.

Large values of $L(n,y,z)$ indicate that the null hypothesis is wrong and
should be rejected. Such large values happen if $n$ is much larger than $y$
but also if $n$ is much smaller. Here, though, we will only reject the null
hypothesis if we have more events in the signal region than are expected
from background, and therefore we reject the null hypothesis if $%
L(n,y,z)>q\chi _{1}^{2}(1-2\alpha )$ and also $n>y$. Here $q\chi _{1}^{2}(p)$
is the $p^{th}$ percentile of a chi-square distribution with one degree of
freedom.

A similar problem, where the background is modeled as a Poisson rather than
a Gaussian, is discussed in much more detail in Rolke, L\'{o}pez \cite{RL}.\
The closely related problem of setting limits was studied in Rolke, L\'{o}%
pez and Conrad \cite{RLC}

\section{Extensions of the Model}

\subsection{Multiple Channels}

In high energy physics we can sometimes make use of multiple channels. We
will discuss the following model: there are $k$ channels and we have $%
N_{i}\sim Pois(r_{i}e_{i}s+b_{i})$, $Y_{i}\sim N(b_{i},\sigma
_{b_{i}}),Z_{i}\sim N(e_{i},\sigma _{e_{i}}),$ $i=1,..,k$, all independent.
The joint density is then found as follows: Let $\mathbf{n}=(n_{1},..,n_{k})$%
, $\mathbf{y}=(y_{1},..,y_{k})$, $\mathbf{z}=(z_{1},..,z_{k})$, $\mathbf{b}%
=(b_{1},..,b_{k})$, $\mathbf{e}=(e_{1},..,e_{k})$, then

\begin{equation*}
\begin{tabular}{l}
$f(\mathbf{n},\mathbf{y,z};s,\mathbf{b,e})=$ \\ 
$\tprod\limits_{i=1}^{k}\frac{\left( r_{i}e_{i}s+b_{i}\right) ^{n_{i}}}{%
n_{i}!}e^{-\left( r_{i}e_{i}s+b_{i}\right) }\frac{1}{\sqrt{2\pi }\sigma
_{b_{i}}}\exp \left( -\frac{1}{2}\frac{(y_{i}-b_{i})^{2}}{\sigma _{b_{i}}^{2}%
}\right) \frac{1}{\sqrt{2\pi }\sigma _{e_{i}}}\exp \left( -\frac{1}{2}\frac{%
(z_{i}-e_{i})^{2}}{\sigma _{e_{i}}^{2}}\right) $%
\end{tabular}%
\end{equation*}%
The log-likelihood function is given by:

\begin{equation*}
\begin{tabular}{l}
$\log Like(s,\mathbf{b,e;n},\mathbf{y,z})=$ \\ 
$\tsum\limits_{i=1}^{k}{\Large [}n_{i}\log \left( r_{i}e_{i}s+b_{i}\right)
-\log (n_{i}!)-(r_{i}e_{i}s+b_{i})-$ \\ 
$\frac{1}{2}\log (2\pi \sigma _{b_{i}}^{2})+\frac{(y_{i}-b_{i})^{2}}{\sigma
_{b_{i}}^{2}}-\frac{1}{2}\log (2\pi \sigma _{e_{i}}^{2})+\frac{%
(z_{i}-e_{i})^{2}}{\sigma _{e_{i}}^{2}}{\Large ]}$%
\end{tabular}%
\end{equation*}

and taking derivatives we find the following system of equations for the
maximum likelihood estimators:

\begin{equation*}
\begin{tabular}{l}
$\tsum\limits_{i=1}^{k}\left( \frac{n_{i}r_{i}e_{i}}{r_{i}e_{i}s+b_{i}}%
-r_{i}e_{i}\right) =0$ \\ 
$\frac{n_{i}}{r_{i}e_{i}s+b_{i}}-1+\frac{y_{i}-b_{i}}{\sigma _{b_{i}}^{2}}=0$
$\ \ i=1,..,k$ \\ 
$\frac{n_{i}r_{i}s}{r_{i}e_{i}s+b_{i}}-r_{i}s+\frac{z_{i}-e_{i}}{\sigma
_{e_{i}}^{2}}=0$ \ $i=1,..,k$%
\end{tabular}%
\end{equation*}%
This system can not be solved analytically but it is fairly easy to do so
numerically, for example with MINUIT. In addition, it can be shown
analytically that the likelihood ratio test statistic depends not on the
absolute values of the efficiencies and branching fractions but only on the
ratios between the values for the different channels.

For the numerator of the likelihood ratio statistic we have $s=0 $ and the
corresponding system has the solutions

\begin{equation*}
\begin{tabular}{l}
$\widetilde{b_{i}}=\frac{1}{2}\left( y_{i}-\sigma _{b_{i}}^{2}+\sqrt{%
(y_{i}-\sigma _{b_{i}}^{2})^{2}+4n_{i}\sigma _{b_{i}}^{2}}\right) $ \ \ $%
i=1,..,k$ \\ 
$\widetilde{e}_{i}=z_{i}$ \ $i=1,..,k$%
\end{tabular}%
\end{equation*}

As above we will claim a discovery only if there is an excess of events in
the signal region. If we denote the test statistic by $L(\mathbf{n},\mathbf{y%
},\mathbf{z})$ this means to reject the null hypothesis of no signal if $L(%
\mathbf{n},\mathbf{y},\mathbf{z})>q\chi _{1}^{2}(1-2\alpha )$ and also $%
\widehat{s}>0$ where $\widehat{s}$ is the maximum likelihood estimator of
the true signal $s$.

\subsection{Extension II: Marked Poisson}

It is sometimes possible to include further information in this model.
Consider the following case: in the $i^{th}$ channel we observe $n_{i}$
events in the signal region and $y_{i}$ events in the background region. We
have an independent measurement $z_{i}$ of the efficiency. Furthermore we
have measurements $x_{ij}$, $j=1,..,n_{i}$ for each event in the signal
region and we know the distributions of these measurements depending on
whether an event is signal or background. This leads to the following
density function:%
\begin{equation*}
\begin{tabular}{l}
$\log Like(s,\mathbf{b,e;n},\mathbf{y,z})=$ \\ 
$\tsum\limits_{i=1}^{k}{\Large [}n_{i}\log \left( r_{i}e_{i}s+b_{i}\right)
-\log (n_{i}!)-(r_{i}e_{i}s+b_{i})-$ \\ 
$\frac{1}{2}\log (2\pi \sigma _{b_{i}}^{2})+\frac{(y_{i}-b_{i})^{2}}{\sigma
_{b_{i}}^{2}}-\frac{1}{2}\log (2\pi \sigma _{e_{i}}^{2})+\frac{%
(z_{i}-e_{i})^{2}}{\sigma _{e_{i}}^{2}}+$ \\ 
$\sum_{j=1}^{n_{i}}\log \left( \frac{r_{i}e_{i}s}{r_{i}e_{i}s+b_{i}}%
f_{i}^{s}(x_{ij})+\frac{b_{i}}{r_{i}e_{i}s+b_{i}}f_{i}^{b}(x_{ij})\right) 
{\Large ]}$%
\end{tabular}%
\end{equation*}%
where $f_{i}^{s}$ and $f_{i}^{b}$ are the densities of the signal and the
background in the $i^{th}$ channel, respectively. In this paper we will
assume that $f_{i}^{s}$ and $f_{i}^{b}$ are fully known but it would be easy
to let them depend on nuisance parameters as well. In some applications
these densities might be estimated from the data, for example using neural
networks. Furthermore, this model allows for a "mixture" case: if in some
channels no measurements $x_{ij}$ are available we only need to set $%
f_{i}^{s}$ and $f_{i}^{b}$ equal to $1$.

The expression above simplifies somewhat if we set $f_{ij}=\frac{%
f_{i}^{b}(x_{ij})}{f_{i}^{s}(x_{ij})}$ and omit any constant terms:%
\begin{equation*}
\begin{tabular}{l}
$\log Like(s,\mathbf{b,e;n,y,z})=$ \\ 
$\sum_{i=1}^{k}\left[ -\left( r_{i}e_{i}s+b_{i}\right) -\frac{1}{2}\frac{%
(y_{i}-b_{i})^{2}}{\sigma _{b_{i}}^{2}}-\frac{1}{2}\frac{(z_{i}-e_{i})^{2}}{%
\sigma _{e_{i}}^{2}}+\sum_{j=1}^{n_{i}}\log \left(
r_{i}e_{i}s+b_{i}f_{ij}\right) \right] $%
\end{tabular}%
\end{equation*}

Finding the maximum likelihood estimators now means solving the following
nonlinear system of $2k+1$ equations:%
\begin{equation*}
\begin{tabular}{l}
$\tsum\limits_{i=1}^{k}r_{i}e_{i}\left( \sum_{j=1}^{n_{i}}\frac{1}{%
r_{i}e_{i}s+f_{ij}b_{i}}-1\right) =0$ \\ 
$\frac{y_{i}-b_{i}}{\sigma _{b_{i}}^{2}}-1+\sum_{j=1}^{n_{i}}\frac{f_{ij}}{%
r_{i}e_{i}s+f_{ij}b_{i}}=0$ $\ \ i=1,..,k$ \\ 
$\frac{z_{i}-e_{i}}{\sigma _{e_{i}}^{2}}-r_{i}s+\sum_{j=1}^{n_{i}}\frac{%
r_{i}s}{r_{i}e_{i}s+f_{ij}b_{i}}=0$ $\ \ i=1,..,k$%
\end{tabular}%
\end{equation*}

Again this system can not be solved analytically. For the numerator of the
likelihood ratio statistic we have $s=0$ and the corresponding system has
the same solutions as the corresponding system in section 4.1. The test is
then again: reject the null hypothesis of no signal if $L(\mathbf{n,y,z,x}%
)>q\chi _{1}^{2}(1-2\alpha )$ and $\widehat{s}>0$.

\section{Performance}

How do the above tests perform? In order to be a proper test they first of
all have to achieve the nominal type I error probability $\alpha $. If they
do, we can then further study their performance by considering their power
function $\beta (s)$ given by%
\begin{equation*}
\beta (s)=P(\text{reject }H_{0}|\text{ true signal rate is }s)
\end{equation*}%
Of course, we have $\beta (0)=\alpha $. $\beta (s)$ gives us the discovery
potential, that is the probability of correctly claiming a discovery if the
true signal rate is $s>0$.

Performance studies for the case of one channel were previously done in
Rolke-Lopez \cite{RL}.

In high energy physics discoveries usually require a very small type I error
probability, often as small as $\alpha =2.87\cdot 10^{-7}$, equivalent to a $%
5\sigma $ event. A straightforward simulation study would therefore need to
do about $10^{9}$ runs. Instead of a simple MC study we will use a technique
called importance sampling to estimate the true type I error probability. It
works as follows. In a straightforward MC study we would generate $N_{i}\sim
Pois(b_{i})$, $Y_{i}\sim N(b_{i},\sigma _{b_{i}})$, $Z_{i}\sim
N(e_{i},\sigma _{e_{i}})$, $X_{ij}\sim F_{i}^{b}$, where $F_{i}^{b}$ is the
distribution of the background events in channel $i$, with $i=1,$ .., $k,$ $%
j=1,$ ..$,$ $N_{i}$. Then we would calculate $L(\mathbf{N,Y,Z,X})$ and find
the percentage of runs where $L(\mathbf{N,Y,Z,X})>q\chi _{1}^{2}(1-2\alpha )$
and $\widehat{s}>0$. At $5\sigma $ though, this will only happen about 1 in
every $3.5$ million runs. So instead we will generate the MC data as if the
true observed signal rate in every channel were $t$, that is we generate $%
N_{i}^{s}\sim Pois(t)$, $N_{i}^{b}\sim Pois(b_{i})$, $Y_{i}\sim
N(b_{i},\sigma _{b_{i}})$, $Z_{i}\sim N(e_{i},\sigma _{e_{i}})$, $X_{ij}\sim
F_{i}^{s}$ for $j=1,..,N_{i}^{s}$ and $X_{ij}\sim F_{i}^{b}$ for $%
j=N_{i}^{s}+1,..,N_{i}^{s}+N_{i}^{b}$ $(=N_{i})$, respectively. For a
suitably chosen $t$, $L(\mathbf{N,Y,Z,X})$ will be of the order of the
critical value reasonably often. We generate $M$ MC samples and find the
true type I error as%
\begin{equation*}
\widehat{\alpha }=\frac{1}{M}\sum_{m=1}^{M}I\left[ L(\mathbf{N,Y,Z,X})>q\chi
_{1}^{2}(1-2\alpha ),\widehat{s}>0\right] w_{m}
\end{equation*}%
where the weights $w_{m}$ are given by the likelihood ratio of the true
density and the one used for the sampling:%
\begin{equation*}
w_{m}=\tprod\limits_{i=1}^{k}\frac{P(N=n_{i}|b_{i})}{P(N=n_{i}|t+b_{i})}%
=e^{kt}\tprod\limits_{i=1}^{k}\left( \frac{b_{i}}{t+b_{i}}\right) ^{n_{i}}
\end{equation*}%
For more on importance sampling see Srinivasan \cite{Srinivasan}.

In figure 1 we have the result of the following study: we use $5$ channels,
the background rates $b$ vary from $2$ to $100$ and are the same in all
channels, $\sigma _{b}=b/15$, $e=0.9$, $\sigma _{e}=0.1$, $r_{i}=0.15$\ in
all channels. As we can see the test achieves the true type I error for all
cases.

Next we will consider what happens when the number of channels grows. In
figure 2 we have $k=1$ to $50$, in all channels $b=25$ with $\sigma _{b}=5/3$%
, $e=0.9$ with $\sigma _{e}=0.09$ and $r=1/k$. Again we achieve the nominal
type I error probability, even for $50$ channels and at $5\sigma $.

In figure 3 we consider the power of the test. There are $10$ channels, each
with a background rate $b=50$, $\sigma _{b}=5$, efficiency $e=0.9$, $\sigma
_{e}=0.09$ and branching ratio $r=0.05$. At the $5\sigma $ level the total
signal rate has to be about $410$ to have a $90\%$ chance of claiming a
discovery.

Now we turn to a study of the marked Poisson case. We will consider two
examples. In the first we have some auxiliary measurement thought to be able
to separate signal and background. The functions $f_{i}^{s}$ and $f_{i}^{b}$
have an (assumed to be known) parameter $\gamma $ and are given by

\begin{equation*}
\begin{array}{l}
f^{s}(x)=\frac{1}{\gamma }e^{\frac{x}{\gamma }},\text{ }1<x<2 \\ 
f^{b}(x)=\frac{1}{\gamma }e^{\frac{3-x}{\gamma }},\text{ }1<x<2%
\end{array}%
\end{equation*}%
For small values of $\gamma $ there is a large distinction between signal
and background, for larger values the separation becomes smaller. Two cases
are shown in figure 4 in the top two panels with two different values of $%
\gamma $ corresponding to strong and almost no separation. $f_{i}^{s}$ is
drawn in dashed lines and $f_{i}^{b}$ in solid lines.

In a different example we use the mass distributions themselves. We assume a
flat background and a Gaussian signal with mean $0.5$ and standard deviation 
$\delta $. Again two cases of different separation between signal and
background are shown in figure 4, in the bottom panels. $f_{i}^{s}$ is drawn
in dashed lines and $f_{i}^{b}$ in solid lines.

We begin as before with a study of the true type I error probability $\alpha 
$. In figure 5 we have $5$ channels. Each channel has a background rate $b$
from $5$ to $50$, $\sigma _{b}=b/5$, an efficiency of $e=0.9$, $\sigma
_{e}=0.1$ and $r=0.15$. The + symbols are for example 1, $\gamma =4$, x for
example 1, $\gamma =0.33$, diamonds for example 2, $\delta =0.25$ and upside
down diamonds for example 2, $\delta =0.05$. For all those cases the method
achieves the true type I error probability $\alpha $.

Finally, in Figure $6$ we present a power study of those same marked Poisson
examples. The signal rate goes from $0$ to $200$. For example 1 we use $%
\gamma =2.0,1.0$ and $0.5$, for example 2 $\delta =0.25$, $0.15$ and $0.05$,
going from a large separation between signal and background by $f^{b}$ and $%
f^{s}$ to almost no separation. Figure $6$ clearly shows how much
improvement is possible by using the extra information contained in $f^{b}$
and $f^{s}$.

The studies here have used reasonable values for the parameters involved.
For example, when using multiple channels, it is reasonable to use channels
for which the product of efficiency times branching fraction is similar. In
our studies these have been set equal. However, an exhaustive performance
study is not possible because of the high dimensionality of the problem.
Nevertheless, we believe that the uniformly excellent performance in studied
cases is an indication that this test will perform very well in a wide range
of cases. In general, though, we would recommend the practitioner to carry
out their own simulation study for their specific problem to insure that the
method also performs well there.

\section{Summary}

We have discussed a hypothesis test for the presence of a signal. We
extended the test to the case of multiple channels as well as the use of
auxiliary measurements using marked Poisson models. Studies of the
performance of the test for typical cases yielded highly satisfactory
results.

\section{Appendix}

\begin{figure}
	\centering
		\includegraphics[width=0.90\textwidth]{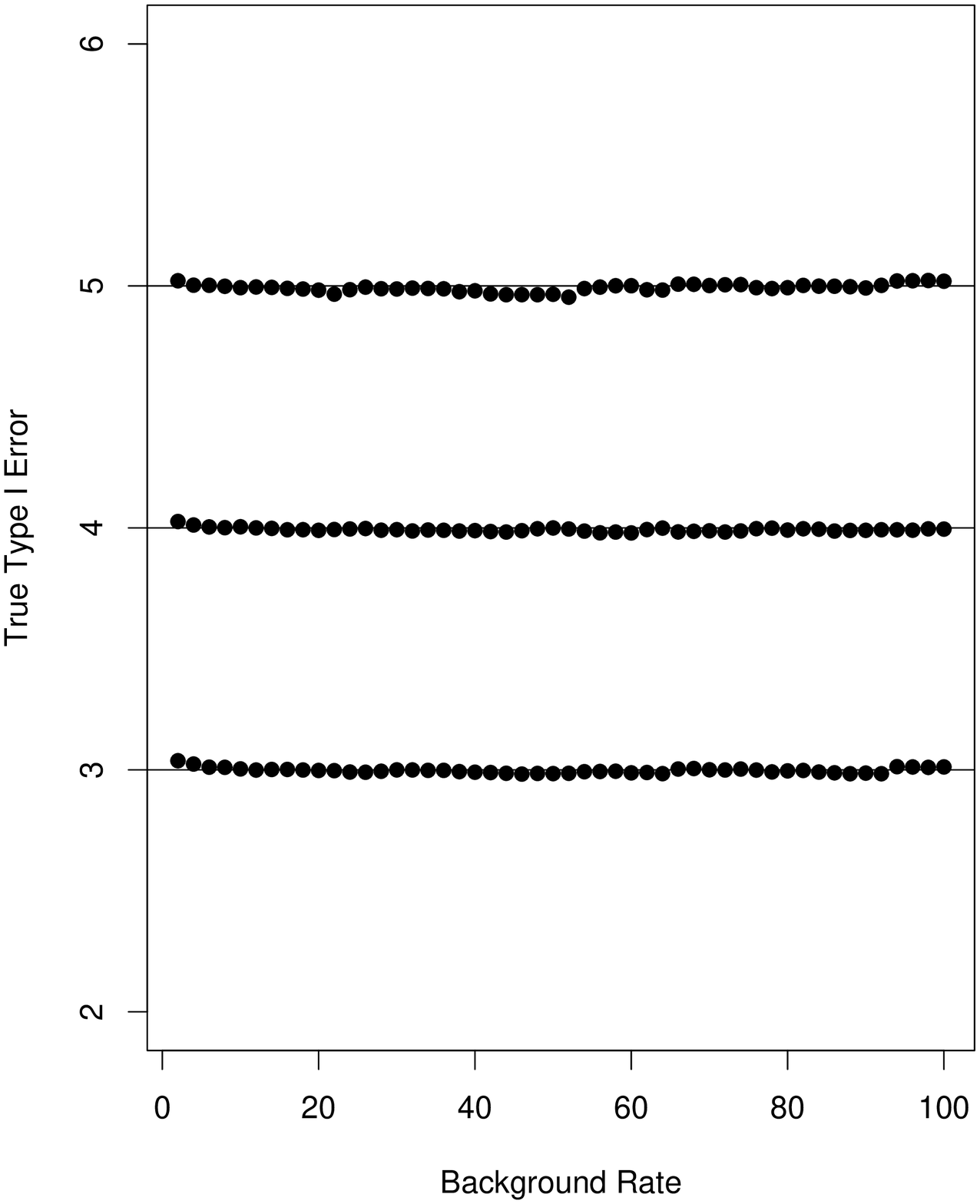}
	\caption{Study of type I error $\protect\alpha $ for case of $5$ channels. The background rates $b$ vary from $2$ to 
$100$ and are the same in all channels. $\protect\sigma _{b}=b/15$, $\protect%
\varepsilon =0.9$, $\protect\sigma _{e}=0.1$, $r_{i}=0.15$\ in all channels.}
	\label{fig:fig1}
\end{figure}

\begin{figure}
	\centering
		\includegraphics[width=0.90\textwidth]{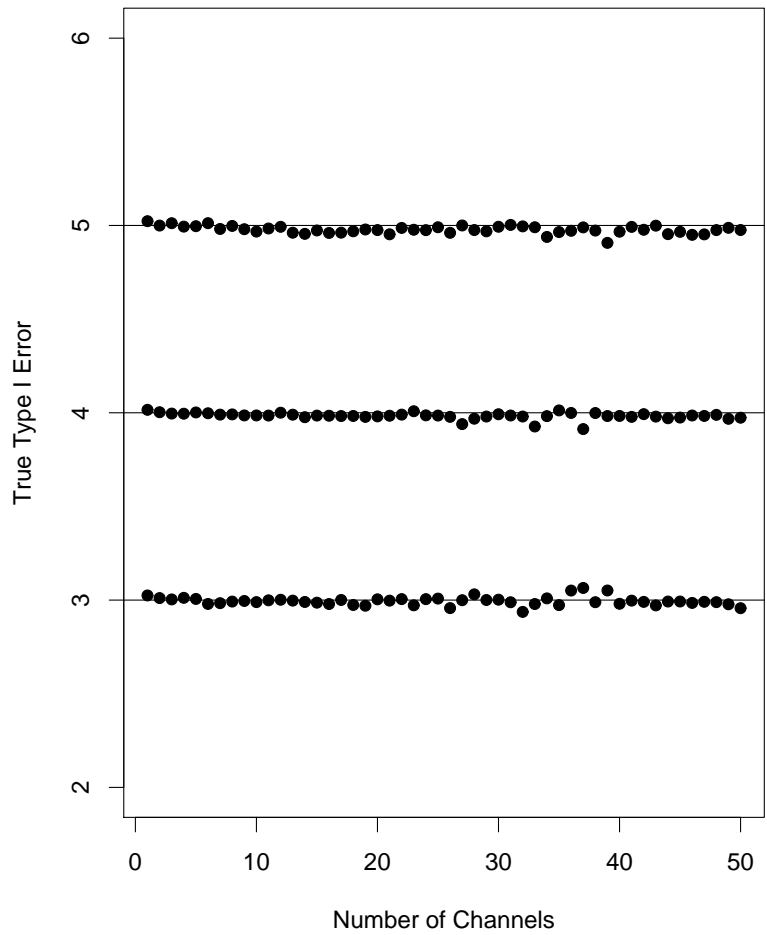}
	\caption{Study of the effect of the number of channels on type I error. There
are $k$ channels ($k=1$ to $50$). In all channels $b=25$ with $\protect%
\sigma _{b}=5/3$, $e=0.9$ with $\protect\sigma _{e}=0.09$ and $r=1/k$.}
	\label{fig:fig2}
\end{figure}

\begin{figure}
	\centering
		\includegraphics[width=0.90\textwidth]{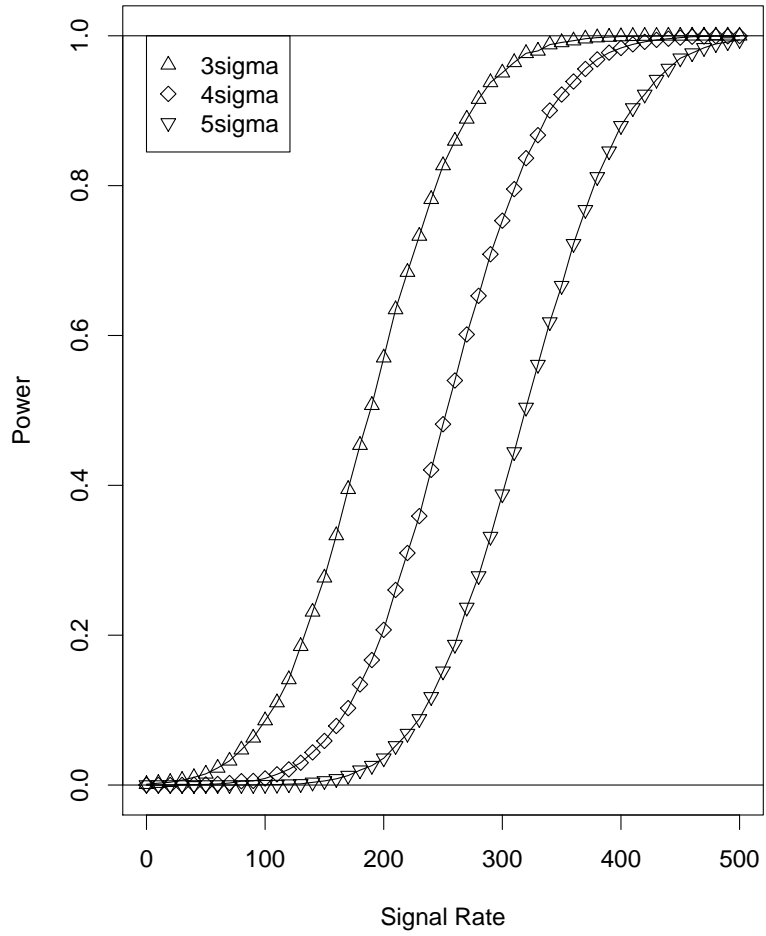}
	\caption{Study of the power of the test. There are $10 $
channels, each with a backgound rate $b=50$, $\protect\sigma _{b}=5$,
efficiency $e=0.9 $, $\protect\sigma _{e}=0.09$ and branching fraction $%
r=0.05$.}
	\label{fig:fig3}
\end{figure}

\begin{figure}
	\centering
		\includegraphics[width=0.90\textwidth]{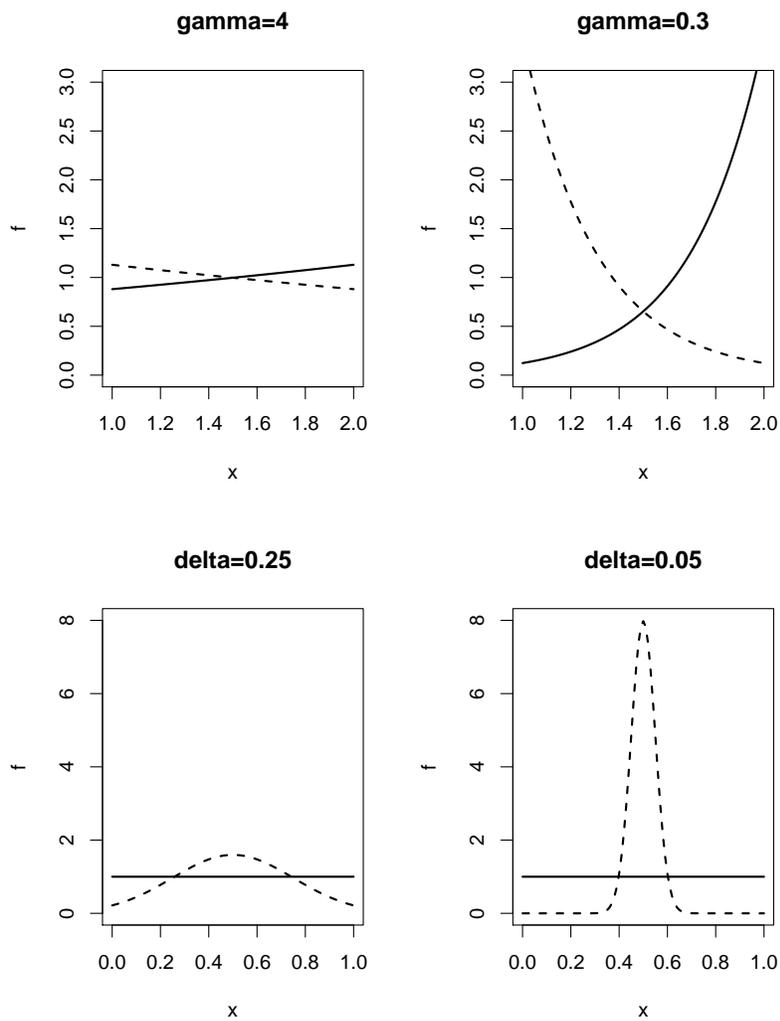}
	\caption{Examples for the two types of distributions used in the
study of the marked Poisson case. In the upper two panels we have an
auxiliary measurement for the events, in the lower two panels the actual
mass distributions are used.}
	\label{fig:fig4}
\end{figure}

\begin{figure}
	\centering
		\includegraphics[width=0.90\textwidth]{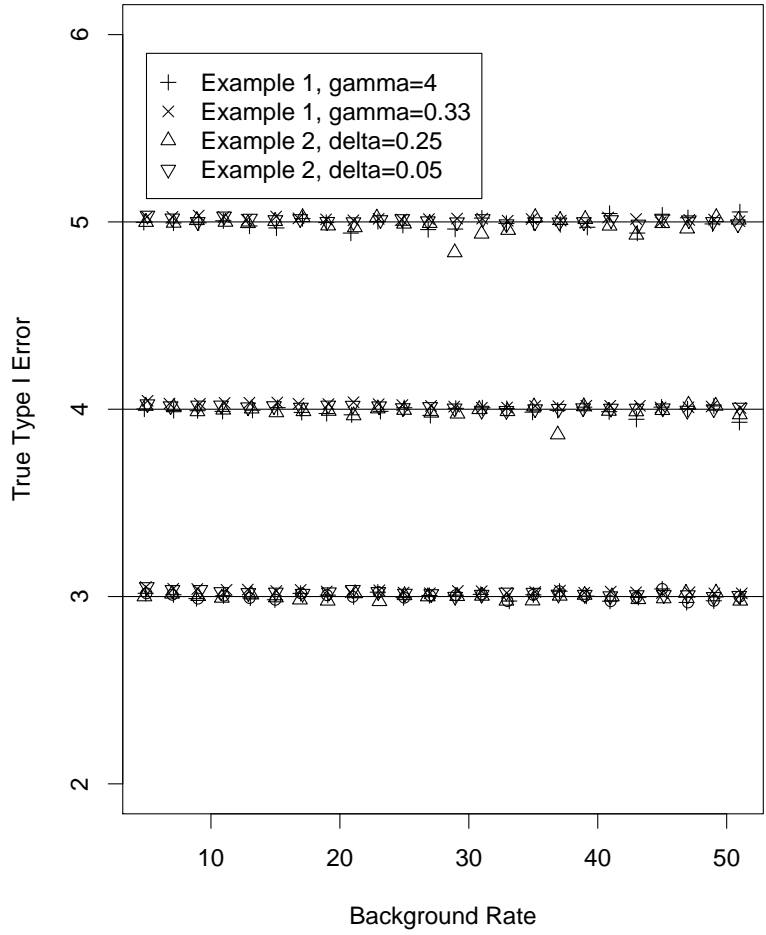}
	\caption{Study of type I error using mulitple
channels and marked Poisson. There are $5$ channels. Each channel has a
background rate $b$ from $5$ to $50$, $\protect\sigma _{b}=b/5$, an
efficiency of $e=0.9$, $\protect\sigma _{e}=0.1$ and $r=0.15$. The + symbols
are for example 1, $\protect\gamma =4.0$, x for example 1, $\protect\gamma %
=0.33$, diamonds for example 2, $\protect\delta =0.25$ and upside down
diamonds for example 2, $\protect\delta =0.05$.}
	\label{fig:fig5}
\end{figure}

\begin{figure}
	\centering
		\includegraphics[width=0.90\textwidth]{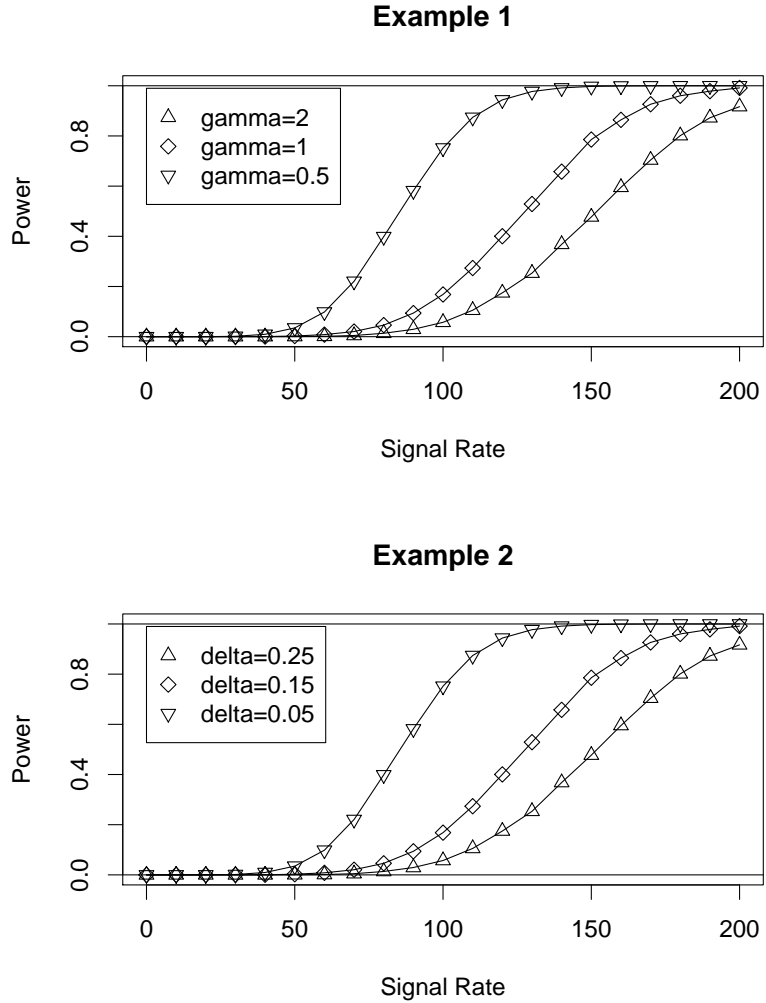}
	\caption{Power study for case of multiple channels and marked Poisson. We use $5$
channels, each channel has a background rate $b=50$, $\protect\sigma _{b}=5$%
, an efficiency of $e=0.9$, $\protect\sigma _{e}=0.1$ and $r=0.15$. The
signal rate goes from $0$ to $200$. For example 1 we use $\protect\gamma =2$%
, $1$, and $0.5$, for example 2 $\protect\delta =0.25$, $0.15$ and $0.05$,
going from a weak separation between signal and backgound by $f^{b}$ and $%
f^{s}$ to a strong separation.}
	\label{fig:fig6}
\end{figure}

\end{document}